\documentclass[aps,pra,twocolumn,superscriptaddress,showpacs,amsmath]{revtex4-1}
\usepackage{graphicx,hyperref}

\newcommand{\ud}{\textrm{d}}
\def\NAT@sort{1}

\begin{document}

\author{Gytis Kulaitis}
\affiliation{Scottish Universities Physics Alliance, School of Physics and Astronomy, University of St Andrews, St Andrews KY16 9SS, United Kingdom}

\author{Frank Kr\"uger}
\affiliation{Scottish Universities Physics Alliance, School of Physics and Astronomy, University of St Andrews, St Andrews KY16 9SS, United Kingdom}

\author{Felix Nissen}
%\email{fbfn2@cam.ac.uk}
\affiliation{Cavendish Laboratory, University of Cambridge, Cambridge CB3 0HE, United Kingdom}

\author{Jonathan Keeling}
\affiliation{Scottish Universities Physics Alliance, School of Physics and Astronomy, University of St Andrews, St Andrews KY16 9SS, United Kingdom}

\title{Disordered driven coupled cavity arrays: Non-equilibrium
  stochastic mean-field theory.}
\begin{abstract}
  We study the interplay of disorder with pumping and decay in coupled
  qubit-cavity arrays, the Jaynes-Cummings-Hubbard model.  We find
  that relatively weak disorder can wash out the bistability present
  in the clean pumped system, and that moreover, the combination of
  disorder in on-site energies and decay can lead to effective phase
  disorder.  To explore these questions, we present a non-equilibrium
  generalization of Stochastic-Mean-Field theory, providing a simple
  tool to address such questions.  This technique is developed for
  rather general forms of light-matter coupling, driving, dissipation,
  and on-site disorder, making it applicable to a wide range of
  systems.
  \end{abstract}

\maketitle

\section{Introduction}
\label{sec:introduction}

Quantum simulation~\cite{Lloyd1996a} concerns how controllable quantum
systems can be used to model particular desirable Hamiltonians, in
order to find the ground state, or other properties, of otherwise
hard-to-simulate problems.  Recently, there has been significant
progress in realizing quantum emulators based on systems including
ultracold atoms~\cite{Bloch2008}, Rydberg atoms~\cite{Weimer2010a},
Trapped Ions \cite{Lanyon2011,Barreiro2011a} or superconducting qubits
in microwave cavities \cite{Koch2009,Houck2012a,Schmidt2012}.  One
approach that has been used recently for cold atoms is to engineer an
effective Hamiltonian in a rotating frame, by using a Raman pumping
scheme~\cite{Dimer:Proposed}. This approach has been used to realize
the superradiance transition in the Dicke model~\cite{Baumann:Dicke}.
In such cases, and more generally for coupled matter-light systems
such as superconducting qubits in microwave cavities, it can be
crucially important to understand the effects of losses and
dissipation.  For example, in the Dicke model, the presence of losses
means the critical behavior~\cite{Nagy2011a,Bhaseen2012,Oztop2012}
becomes classical~\cite{Torre2012a} due to the effective temperature
introduced by losses.  Similar issues can generally be expected to
occur in any open driven system, and this therefore may have
consequences across the range of experimental systems considered as
potential quantum emulators, and in particular, for coupled
light--matter systems \cite{Hartmann2008}.

Another model where quantum simulation has been
explored~\cite{Trotzky2010} is the disordered Bose-Hubbard model
(BHM)~\cite{Fisher:Bosonloc}.  This model consists of bosonic
particles hopping between sites with repulsion between particles on
the same lattice site.  This model can be simulated with ultracold
bosonic atoms, introducing disorder in a highly controlled manner by
superimposing a fine-grained optical speckle potential with a periodic
optical lattice \cite{Lye+05,Clement+08,Billy+08,White+09}.  In the
presence of weak disorder in the on-site energies, three possible
ground states exist at zero temperature: a superfluid phase and two
insulating phases.  The two insulating phases are the incompressible
Mott Insulator and the compressible Bose glass. In the Mott Insulator,
the particles are localized because of strong local repulsions, in the
Bose glass particles are localized because of the disorder potential.
Despite the long history of the BHM, it is only recently that several
aspects of this model have been fully understood, such as confirmation
that the Bose-glass phase always intervenes between Mott insulator and
superfluid~\cite{Gurarie2009}, and the distinction between the Mott
insulator and Bose glass regarding whether fluctuations are
self-averaging \cite{Kruger2011a}.  Even if quantum simulation of such
a model with an effective Hamiltonian in a dissipative system can only
model the finite temperature case, this may of itself be enough to
answer questions such as whether the finite temperature insulating
phase is self-averaging.  However, as we will discuss further below,
disorder and non-equilibrium effects can conspire to significantly
change the behavior (and universality class) of the model system.

The Jaynes-Cummings-Hubbard model (JCHM)~\cite{Greentree2006} is very
closely related~\cite{Koch2009} to the BHM, and can be more directly
realized in coupled cavity
arrays~\cite{Hartmann2008,Koch2009,hur10,Houck2012a,Schmidt2012}.  This
model naturally describes superconducting qubits in microwave
cavities.  The JCHM consists of photons coupled to two-level systems,
considering photons confined in an array of coupled cavities, with
weak hopping between different cavities.  The JCHM requires that only
the energy preserving term $(a \sigma^+ + a^\dagger \sigma^-)$ in
qubit-cavity coupling is important. When counter-rotating terms $(a
\sigma^- + a^\dagger \sigma^+)$ are also important the model is known
as the Rabi-Hubbard model, and the symmetry of the Hamiltonian is
lowered from $U(1)$ to $Z_2$ and the phase diagram significantly
changes \cite{Schiro2012}.  In the case of the JCHM, previous work has
shown how in equilibrium, including on-site disorder leads to
behavior very similar to the BHM~\cite{Mascarenhas2012a}, as may be
expected from the symmetries of the problem~\cite{Koch2009}.

In this paper we study the non-equilibrium JCHM in the presence of
disorder.  We focus on the simplest possible form of pumping and
decay, i.e.  uniform coherent pumping, as has previously been studied
in the clean limit~\cite{Nissen2012,Grujic2012,Grujic2012a}.  In this case, all
symmetries are broken by the pumping, and no phase transitions are
expected.  Nonetheless, the behavior we observe and discuss for this
case clearly shows how new physics would also arise with other forms
of pumping which need not break the symmetries of the model.  In
particular, we see that pumping and dissipation can transform on-site
energy disorder into phase disorder, destroying long range order in
the superfluid phase.  In addition, we explore the fate of the
bistability seen in the clean non-equilibrium JCHM~\cite{Nissen2012}.

To explore these questions, we use a generalized ``Stochastic''
Mean-Field theory~\cite{Weber2006,Bissbort2009a,Bissbort2010}, which
involves self-consistency equations for the probability distribution
of local order parameters.  We extend this approach to apply to open
quantum systems.  Such an approach is approximate, and only becomes
well controlled at high coordination number (i.e. in high dimensions).
Nonetheless, it provides a simple tool to effectively explore the
interplay of disorder and pumping, and see whether effective
Hamiltonians for open systems could \emph{in principle} be used to
simulate disordered quantum systems.

The remainder of this paper is arranged as follows. In 
Sec.~\ref{sec:stochatic-mean-field} we generalize stochastic mean-field
theory (SMFT) to treat disorder in open quantum systems. The technique is 
introduced for rather general forms of pumping, decay, and on-site
disorder. As an example, we apply the SMFT to the dissipative JCHM.
In Sec.~\ref{sec:results}, we first briefly summarize the behavior of the 
JCHM in the absence of disorder and then discuss the effects of 
on-site disorder in the excitation energies of the two-level systems.

\section{Stochastic Mean Field Theory of open systems}
\label{sec:stochatic-mean-field}

This section briefly summarizes the SMFT approach as
applied to the non-equilibrium problem.  The equilibrium SMFT was 
introduced in the context of disordered
antiferromagnets \cite{Weber2006} and later applied to the
BHM~\cite{Bissbort2009a,Bissbort2010} and has more recently been
applied to the JCHM \cite{Mascarenhas2012a}.  We present the following
discussion for a general coupled cavity array problem, and specialize
to the JCHM in section~\ref{sec:results}.

We consider an array of cavities with coordination number $z$ and hopping
$J/z$ of photons between neighboring cavities, given by the Hamiltonian
\begin{equation}
\label{eq:1}
H = \sum_i h_i - \frac{J}{z} \sum_{\langle ij \rangle} a^{\dagger}_i a_j,
\end{equation}
where $a^\dagger_i$ ($a_i$) creates (annihilates) a photon on the
$i$th cavity. The on-site Hamiltonian
$h_i=h(a_i,X_i^{(\alpha)},\epsilon_i)$ for the individual cavities can
be completely general at this point. The operators $X^{(\alpha)}$ act
on the Hilbert space of the possible quantum states of the matter
contained in the cavities. In the simplest cases, including the JCHM
and the Rabi-Hubbard model, this will be a two-level system and the
$X$ operators are spin-1/2 operators. The on-site Hamiltonian will
contain a coupling between the photons and the matter degrees of
freedom as well as any coherent pumping terms.

We further introduce on-site disorder $\epsilon_i$ which can couple
either to the photon energy or to the matter in the cavity. The
disorder follows a probability distribution $p(\epsilon)$ and is
assumed to be uncorrelated between different cavities.  For such
on-site disorder the method is as developed in
Refs.~\cite{Bissbort2009a,Bissbort2010}.  If one instead considered
disorder in the hopping between sites, the problem is analogous to
that originally considered in Ref.~\cite{Weber2006}.

Dissipation is included on the level of a master equation for the time
evolution of the density operator,
\begin{equation}
\label{eq:2}
 \frac{\ud\rho}{\ud t} = -i \left[H,\rho \right]+ \sum_i \left\{\frac{\kappa}{2} \mathcal{L}[a_i]+\sum_\alpha \frac{\gamma_{\alpha}}{2} \mathcal{L}[X^{(\alpha)}_i]\right\},
\end{equation}
where $\mathcal{L}[X] = 2 X \rho X^\dagger - \left\{X^\dagger X, \rho\right\}$ denote the standard Lindblad operators.

The basic idea of SMFT is to consider a self-consistency condition for
the probability distribution $P(\psi)$ of on-site coherent fields
$\psi=\langle a \rangle$.  From this one may find the distribution of
sums of fields from neighboring sites:
\begin{equation}
  \label{eq:3}
  Q(\phi) = \int \prod \ud \psi_i \delta(\phi - \sum_i \psi_i) P(\psi_i)
\end{equation}
where the product and sum run over the $z$ nearest neighbors.  The relation 
between $P$ and $Q$ simplifies in Fourier space
\begin{equation}
  \label{eq:4}
  \tilde{Q}(\xi) = \int \ud\phi Q(\phi) e^{i \xi \phi}, \qquad
  {Q}(\phi) = \int \frac{\ud\xi}{2\pi} \tilde{Q}(\xi) e^{-i \xi \phi}
\end{equation}
for the $Q$ distribution. Using the convolution theorem we obtain 
$\tilde{Q}(\xi) = \tilde{P}(\xi)^z$.

Given the distribution of fields from neighboring sites, the
self-consistency condition comes from assuming that this distribution
of fields is uncorrelated with the site energies,  and so one may write:
\begin{equation}
  \label{eq:5}
  P(\psi) = \int \ud \phi \int \ud \epsilon
  Q(\phi) p(\epsilon) \delta(\psi - \lambda(\phi,\epsilon)).
\end{equation}
Here $\lambda(\epsilon,\phi)$ gives the expectation for $\psi$
corresponding to a field $\phi$ from the neighbors, and on-site
energy $\epsilon$.  In our case this corresponds to finding the
steady-state on-site density matrix from,
\begin{subequations}
\label{eq:6}
\begin{eqnarray}
 \frac{\ud \rho_i}{\ud t} & = & - i \left[ h^{\text{eff}}_i, \rho_i \right]+
 \frac{\kappa}{2} \mathcal{L}[a_i]+\sum_\alpha \frac{\gamma_{\alpha}}{2} \mathcal{L}[X^{(\alpha)}_i]\\
 h^{\text{eff}}_i & = & h_i-\frac{J}{z}
 \left( \phi a^\dagger_i + \phi^\ast a^{}_i \right)
\end{eqnarray}
\end{subequations}
and determining the expectation value $\lambda(\phi,\epsilon_i) = \mathrm{Tr}(a_i
\rho_i)$.  In steady state, $\ud \rho_i/ \ud t=0$, the master equation
for the onsite density operator (\ref{eq:6}) turns into a set of
coupled linear equations for the matrix elements
$(\rho_i)_{mn}=\langle m|\rho_i|n\rangle$ with respect to a basis of
the product Hilbert space of the matter and photon systems. While the
former is usually finite, we truncate the bosonic Hilbert space at a
certain maximum number of photons per cavity.

As noted above, even for real $\phi$, the values of $\psi$ will be
complex.  This means it is necessary to allow for the distributions
$P$ and $Q$ to extend over complex fields.  Convolution of
two-dimensional distributions follows as before, but with $\phi \to
(\phi^\prime, \phi^{\prime\prime})$ in order to use the convolution
theorem.  In practice, we find it most efficient to pre-calculate an
interpolated approximation to $\lambda(\phi,\epsilon)$, and then iteratively update $P(\psi), Q(\phi)$ until the
distribution converges.  One may also note that it is not guaranteed
that the above iteration procedure should converge, nor that it should
converge to an unique solution --- as discussed below, the mean field
decoupling introduces the possibility of multistability.  However, in
cases where it does converge, the solution found can be regarded as an
approximate description of a possible asymptotic state of the system.
When multiple solutions exist, further work is required to determine
which solution is reached from given initial conditions, and the rate
of tunnelling events that may switch between solutions. This is
discussed further below in Sec.~\ref{sec:summary-clean-jchm}.  In the
cases presented in this paper, only one asymptotic state was found.

\section{Application to the pumped dissipative JCHM}
\label{sec:results}

As a simple application of the above technique, and the simplest kind
of pumped-dissipative array, we consider here the coherently pumped
Jaynes-Cummings-Hubbard model, as studied previously
in~\cite{Nissen2012,Grujic2012,Grujic2012a}.  In terms of the general lattice
problem described in Eq.~(\ref{eq:1}), the Jaynes-Cummings-Hubbard
model that we consider has an on-site Hamiltonian:
\begin{multline}
  \label{eq:7}
  h_i  =   J a_i^\dagger a_i +  \frac{\epsilon_i}{2}  \sigma_i^z
  + g(\sigma_i^+ a_i + \text{H.c.})\\
  + f (a_i e^{i \omega_p t} + \text{H.c.})
\end{multline}
 where $a^\dagger_i$ creates a photon in the $i$th cavity and the spin-1/2 
operators $\sigma_i^+$, $\sigma_i^-$ describe transitions of the state of the 
two-level (artificial) atom on site $i$.
 $f$ denotes the strength of the pumping at frequency $\omega_p$.
 The cavity photon energy
$J$ is chosen so that for $g=0$, the bottom of the photon dispersion
is at zero energy. 
Disorder is introduced by considering a Gaussian distribution of $\epsilon_i$,
of width $\sigma_\epsilon$ and we take the mean value $\bar{\epsilon}=0$, so
that the mean detuning is as in~\cite{Nissen2012}.  
Further, we consider loss 
terms of the form $\sum_i \left\{(\kappa/2) \mathcal{L}[a_i] +
  (\gamma/2) \mathcal{L}[\sigma_i^-]\right\}$.  The problem can be
trivially made time-independent by the Unitary transform $a \to a
e^{-i \omega_p t}, \sigma^- \to \sigma^- e^{-i \omega_p t}$. 

Other than the coherent pumping term, the problem we consider has a
$U(1)$ symmetry, and this can be used to simplify the pre-calculation
of $\lambda(\phi,\epsilon)$ as discussed in the previous section.
The effective on-site problem of the JCHM has a Master equation with
\begin{multline}
    h^{\text{eff}}_i  =   
    \left(J-\omega_p\right) a_i^\dagger a_i +
    \frac{\epsilon_i-\omega_p}{2}  \sigma_i^z
    + g(\sigma_i^+ a_i + \text{H.c.})\\
    + \left[
      \left(f-\frac{J \phi^\ast}{z}\right) a_i  + \text{H.c.}
    \right].
\end{multline}
One may then write the steady state expectation $\text{Tr}(\rho a_i)$ arising
from this effective Hamiltonian along with the Lindblad terms in the form:
\begin{equation}
  \label{eq:12}
  \text{Tr}( a \rho) = \lambda\left(f^\text{eff} \equiv f-\frac{J \phi}{z}, \epsilon\right),
\end{equation}
where the last line of Eq.~8 can be written
as $\ldots [(f^\text{eff})^\ast a_i + f^\text{eff} a_i^\dagger]$,
combining both the explicit pump and the field coming from the
neighboring cavities into $f^\text{eff}$. 
 The advantage of writing the
expression in this form is that one may note that
$\lambda(f^{\text{eff}}, \epsilon_i) =
(f^{\text{eff}}/|f^{\text{eff}}|)
\lambda(|f^{\text{eff}}|,\epsilon_i)$, i.e. the phase of the input and
output are directly related, although \emph{not} equal.

\subsection{Summary of clean JCHM}
\label{sec:summary-clean-jchm}

For comparison, we briefly summarize here the behavior in the absence
of disorder.  In the absence of hopping, the problem is identical to
that studied by \citet{Bishop2008}: The coupled qubit-cavity system
has an anharmonic polariton spectrum, and so at low pumping, one can
consider the response to pumping an effective two-level system.  If
one considers the coherent field amplitude $|\langle a \rangle|$ as a
function of pump frequency $\omega_p$ then at weak pumping there is a
standard Lorentzian response, while at higher power, power
broadening~\cite{Walls:Book} leads to a reduction of the coherent
field amplitude near resonance, i.e. there is an anti-resonance
feature.  Turning on hopping, the location of the anti-resonance
shifts away from the low-power resonance.  Eventually it shifts so far
that the coherent field amplitude vs pump frequency develops a jump
and an associated bistability.  Such bistability is analogous to that
known in the Dicke model when driving above resonance, where
nonlinearity can blueshift the polariton frequency into resonance.

Let us note at this stage that although the existence of bistability
is due to the mean-field decoupling, its presence is indicative of
physically meaningful bimodal distributions in the true density matrix
\cite{Drummond1980,lugiato84}.  The equation of motion for the
full-system density matrix is linear, and so either has a unique
steady state, or a degenerate subspace of steady states.  The
mean-field decoupling instead produces a nonlinear equation for the
single-site density matrix, which may have multiple distinct solutions
--- these distinct solutions can thus describe bistability.  Where
mean-field theory would predict bistability, the full density matrix
would generally have a configuration with a significant weight near
both of these mean-field solutions, but with a fixed ratio between
their weights and a tail of finite probability states that connect
these.  Both the ratio of weights and the existence of the
intermediate states cannot be found by mean field theories, and
require consideration of fluctuations, and specifically instanton and
soliton corrections that would describe tunneling between different
mean-field configurations \cite{kamenev11}.  It is however worth
noting that all these statements relate to the ensemble averaged
steady state of the system.  If a system is prepared near to one of
the two bistable states, the subsequent dynamics will initially remain
near that configuration until a tunneling event causes a transition to
the other state.  Such tunneling (quantum, thermal or induced by
external noise) can cause transitions in both directions, and
eventually produces a fixed ratio between the two parts of the bimodal
distribution.

Since the spacing of energy levels of the JCHM is anharmonic, in the
limit of relatively weak hopping, the problem can be understood
quantitatively by restricting the on-site problem to a reduced Hilbert
space of $0,1$ excitations.  As discussed
in~\cite{Bishop2008,Nissen2012}, this is valid as long as other
excitations are sufficiently far from resonance, $ U_{\text{eff}} \gg
f$ where $U_{\text{eff}}$ is an effective anharmonicity (which
vanishes for large hopping).  This reduces the problem to:
\begin{equation}
  \label{eq:8}
  H_{\text{eff}} = 
  \sum_i \left( \frac{\eta}{2} \tau_i^z +     \tilde{f} \tau_i^x \right)
    - \frac{\tilde{J}}{z} \sum_{\langle ij \rangle} \tau_i^+ \tau_j^-
\end{equation}
where $\tau_i^\alpha$ are Pauli matrices in the reduced Hilbert space
and the effective parameters are $\eta =(J+\epsilon -
\sqrt{(J-\epsilon)^2 + 4 g^2})/2 - \omega_p$, $\tilde{J}=J \sin^2
\theta$, and $\tilde{f}=-f \sin\theta$ with $\tan(2 \theta) =
2g/(J-\epsilon)$.    Losses are described by $\sum_i (\tilde{\kappa}/2)
\mathcal{L}[\tau^-_i]$ with $\tilde \kappa = \kappa \sin^2 \theta +
\gamma \cos^2 \theta$. Since $J \ll g$ is assumed one may further
approximate $\eta \simeq -g + (J+\epsilon)/2 - \omega_p, \theta \simeq
\pi/4$. The steady state of this problem can be
reduced to coupled equations for the coherent field amplitude $\psi =
\langle \tau^- \rangle$, an effective detuning $\Delta = \eta +
\tilde{J}(2n-1)$, and the excited state population $n=\langle 1
+\tau^z \rangle/2$,
\begin{equation}
  \label{eq:9}
  \psi = \frac{ \tilde{f} ( \Delta  -i \tilde{\kappa}/2 )}{
     \Delta^2 + (\tilde{\kappa}/2)^2 + 2 \tilde{f}^2},
  \quad
  n= \frac{ \tilde{f}^2}{ \Delta^2 + (\tilde{\kappa}/2)^2 + 2 \tilde{f}^2}.
\end{equation}
One may thus see that for $\tilde{f}\ll \tilde{\kappa}$ one has
resonance at $ \eta = \tilde{J} = J/2$ giving $\omega_p \simeq -g $.
In contrast, for larger $\tilde{f}$ one has $n \to 1/2$ and the center
of the anti-resonance is at $\eta =0$, i.e. $\omega_p = -g + J/2$.
Such behavior is already clear in the clean limit shown in
Fig.~\ref{fig:modp-mono}, even with $f=\kappa=\gamma$.

For large enough $J$, there are multiple solutions of the above
equations.  Equivalently this means $\eta$ is a non-monotonic function
of $\Delta$, and so by writing:
\begin{equation}
  \label{eq:10}
  \eta = \Delta + \tilde{J} - \tilde{J} \frac{ 2\tilde{f}^2}{ \Delta^2 + (\tilde{\kappa}/2)^2 + 2 \tilde{f}^2}
\end{equation}
one can find the critical value of $\tilde{J}$ for bistability by seeking
the solution of $d \eta/d \Delta=0=d^2 \eta/d \Delta^2$.  This yields:
\begin{equation}
  \label{eq:11}
  \tilde{J}_c = \frac{4}{\tilde{f}^2} \left( \frac{2 \tilde{f}^2 + (\tilde{\kappa}/2)^2}{3} \right)^{3/2}
\end{equation}
For $\tilde{J}>\tilde{J}_c$, there is a range of $\eta$ (i.e. pump
frequencies) for which $\Delta(\eta)$ and thus $\psi(\eta)$ are
multi-valued and so describe bistability.

\subsection{Effects of disorder}
\label{sec:effects-disorder-at}

\begin{figure}[htpb]
  \centering
  \includegraphics[width=\linewidth]{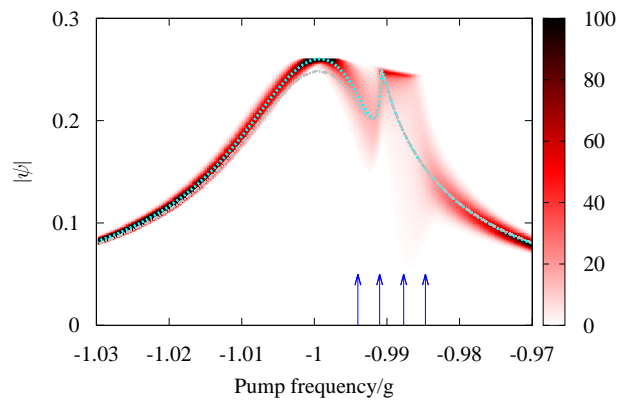}
  \caption{(Color online) Probability distribution of $|\psi|$ as a function of pump
    frequency.  The dashed (cyan) line indicates the value of $|\psi|$
    in the clean limit, and the colormap shows the probability
    distribution of $|\psi|$ for Gaussian disorder of variance
    $\sigma_\epsilon/g=0.002g$.  The dash-dotted (gray) line indicates the
    approximate solution to the clean case given above in
    Eq.~(\ref{eq:6}).  Blue arrows mark the values of pump frequency
    at which the full probability distribution of complex $\psi$ is
    shown in Fig.~\ref{fig:pset_mono}.  Other parameters are
    $f=\kappa=\gamma=0.005g, J/g=0.020$ and a geometry with $z=2$ is
    assumed. %NB.  J=2*t, because code has z*t as offset, and no 1/z.
  }
  \label{fig:modp-mono}
\end{figure}

\begin{figure}[htpb]
  \centering
  \includegraphics[width=\linewidth]{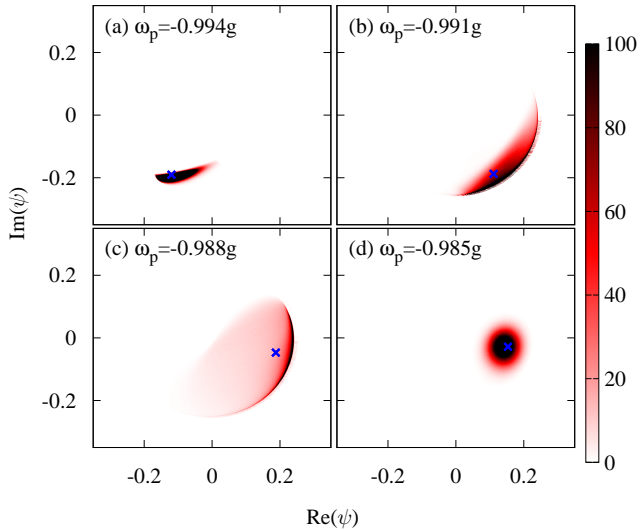}
  \caption{(Color online) Probability distribution of the complex observable $\psi$
    at four different pump frequencies for $J<J_c$.  The blue crosses
    indicate the value of $\psi$ found for the clean case, and the
    colormap shows the probability distribution with
    $\sigma_\epsilon/g=0.002$.  All parameters are as in
    Fig.~\ref{fig:modp-mono}}
\label{fig:pset_mono}
\end{figure}

We consider first the effects of disorder when $J \lesssim J_c$, so
there is no bistability, but a strong distortion compared to $J=0$.
The probability distribution of the amplitude of the coherent field
strength, $P(|\psi|)$ in this case is shown in
Fig.~\ref{fig:modp-mono}, and cross sections, showing the full
probability distribution $P(\psi)$ over the complex plane are given in
Fig.~\ref{fig:pset_mono}.  We consider here parameters as discussed
in~\cite{Nissen2012} for ease of comparison.  For the inclusion of
disorder, one may note that a typical scale of disorder in recent
experiments \cite{Underwood2012} is $\sigma_\epsilon \sim 1$ MHz,
corresponding to $0.002 \lesssim \sigma_\epsilon/g \lesssim 0.005$.  We show
results for $\sigma_\epsilon/g \simeq 0.002$; larger disorders show
very similar behavior.

For most pump frequencies, disorder has a relatively weak effect, but
near the anti-resonance feature it causes a much larger effect.  This
can easily be understood from the discussion of the clean case above:
in this regime $\omega_p \simeq -g + J/2$, and the effective detuning
$\Delta_i \simeq \eta_i \simeq -g + (J+\epsilon_i)/2 - \omega_p \simeq
\epsilon_i/2$.  Thus, near the antiresonance, the variance of $\Delta$
is large compared to its mean value.  Since the variance of disorder
is of the same order as the linewidth $\kappa$, one finds in this
regime that the phase of the on-site order parameter can vary
significantly.  This is clearly seen in Fig.~\ref{fig:pset_mono}(b).
In contrast, away from this point, the mean value of $\Delta$ is much
larger than its variance, and so disorder has only a weak effect on
the phase and amplitude, hence the clean results are recovered.

As one continues to increase the pump frequency above the anti-resonance,
the field amplitude remains notably higher than in the clean case, and
(as seen in Fig.~\ref{fig:pset_mono}(c)) the phase distribution
remains broad.  The increased amplitude can be clearly understood as
an effect of the phase distribution: increasing the phase distribution
means the convolution distribution $Q(\phi)$ moves toward smaller
$\phi$.  Since the field seen by a given site is given by
$f^{\text{eff}} = f - J \phi/z$, and since $\textrm{Re}(\phi)>0$, reducing
$|\phi|$ increases the effective driving, and thus increases the
amplitude.

The phase spreading seen here signifies an important distinction
between the thermal and the non-equilibrium disordered problem.  In
the thermal case, a real distribution of $\psi$ is stable, but in the
non-equilibrium case there is always a distribution of phase, and near
resonance, this becomes particularly notable.  In the current case,
phase symmetry is broken by the external pump.  However, for
incoherent pumping, phase symmetry is not broken.  The presence of
phase spreading then means that following \citet{Imry1975}, no
spontaneous phase symmetry breaking is possible in $d<4$.  This is
quite different from the equilibrium JCHM where a superfluid
(superradiant) state with phase symmetry breaking is expected in
$d>2$.  A similar observation has recently been made for the
disordered polariton condensate \cite{eastham2013}.

\begin{figure}[htpb]
  \centering
  \includegraphics[width=\linewidth]{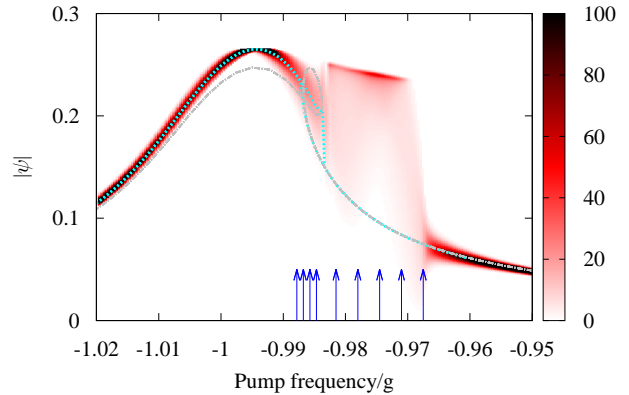}
  \caption{(Color online) Probability distribution of $|\psi|$ as a function of pump
    frequency for $J>J_c$.  Lines and parameters are as for
    Fig.~\ref{fig:modp-mono}, except $J/g=0.04$ in this case. }
  \label{fig:modp-bi}
\end{figure}

\begin{figure}[htpb]
  \centering
  \includegraphics[width=\linewidth]{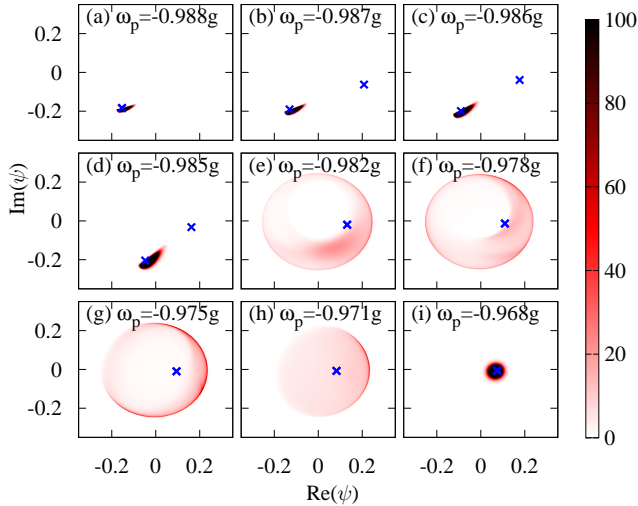}
  \caption{(Color online) Probability distribution of the complex observable $\psi$
    at nine different pump frequencies for $J>J_c$. All parameters are
    as in Fig.~\ref{fig:modp-bi}}
\label{fig:pset_bi}
\end{figure}

As discussed above, in the clean case, for $J>J_c$ bistability occurs
because of the multivalued nature of $\Delta(\eta)$.  However, the
range of detunings where this occurs is the same range where strong
phase spreading was seen, and thus disorder strongly affects the
behavior in exactly this region.  Thus, as seen in
Fig.~\ref{fig:modp-bi}, the disordered case with a typical disorder
strength $\sigma_{\epsilon}/g=0.002$ does not show any bistability,
with no weight near the new clean solution which appears as $\omega_p$
is increased. The absence of bistability is revealed by noting that
the same steady state is found independent of starting distribution of
$P(\psi)$; in the current case this was tested by comparing a
``sweep'' of slowly increasing or decreasing pump frequency, both
cases lead to identical results.  Since a unique solution exists in
these cases, this may be taken as an approximation of the true
asymptotic state of the disordered system.  One may also note in
Figs.~\ref{fig:pset_bi}(e-h) that even once the low $\omega_p$
solution vanishes, weight is not concentrated near the high $\omega_p$
solution until significantly above the antiresonance frequency, as
there is a strong effect of the phase distribution.

\section{Conclusions and open questions}
\label{sec:concl-open-quest}

We have presented a non-equilibrium extension of stochastic mean-field
theory, applicable to problems of coupled cavities with rather general
forms of driving and dissipation.  Using this approach we studied the
effect of disorder on the driven dissipative JCHM.  Near the anti-resonance,
disorder introduces significant phase spreading, which in turn increases
the coherent field amplitude over a range of pump frequencies above
the anti-resonance.  

The results presented above for the dissipative driven JCHM clearly
demonstrate that the combination of open quantum systems with disorder
can lead to behavior that is not seen with only one of these two
ingredients in isolation.  Such behavior prompts an important
question regarding whether dissipative coupled matter-light systems
could ever be used as ``quantum simulators'' of disordered models.  At
the same time, it indicates that there are open questions as to what
the phase diagram of incoherently pumped disordered dissipative
systems may be.  In some cases, such as the Rabi-Hubbard
model~\cite{Schiro2012}, only discrete symmetries exist, and so the
effects of disorder should not destroy the symmetry-broken phase, and
the behavior may be equivalent to that of the site-disordered
transverse field Ising model.  However, for cases with continuous
symmetry it is unclear whether any phase boundaries exist, since
neither the superfluid nor Mott-insulating states survive the effects
of dissipation.  Such questions can be in part addressed by the SMFT
approach described here.

Another challenge is to go beyond the Stochastic-Mean-Field limit
presented here, and produce alternate methods to treat open disordered
lattice problems.  As with all mean-field approaches, SMFT neglects
quantum correlations between different sites; an assumption only valid
in the limit of high coordination.  In addition, SMFT makes a second
assumption, that there is no correlation between the on-site energy
and the field distribution seen.  Such an assumption implies
self-averaging, while it is known that self-averaging breaks down in
the equilibrium Bose glass \cite{Kruger2011a}.  An alternative
approach that may circumvent this is to consider extensions of the
cavity method e.g.~\cite{Semerjian2009}.  For the purpose of
understanding the behavior of currently achievable experiments
\cite{Houck2012a}, finite size simulations of the mean-field
\cite{Nissen2012} or beyond-mean-field \cite{Grujic2012,Grujic2012a} dynamics
 may be more appropriate.  However, a full understanding of the behavior
of such dissipative models may well depend on rare events, not
captured in finite size simulations, so methods such as that presented
here may play an important role.

In conclusion, the combination of dissipation and disorder can lead to
types of behavior in coupled cavity arrays that cannot be seen in
either the clean non-equilibrium system, or disordered equilibrium
case.  This suggests that such cavity arrays may not be
appropriate as quantum simulators to understand equilibrium disordered
problems.  Stochastic mean-field theory can provide a simple route to
address some classes of system, but leads to questions that require
more sophisticated approaches to non-equilibrium disordered problems.

\acknowledgements{GK acknowledges support from the Nuffield trust, and
  University of St Andrews URIP program.  FBFN acknowledges support
  from EPSRC.  JK and FBFN acknowledge discussions with H. T\"ureci and
  S. Schmidt.  JK acknowledges useful discussions with M. Schiro,
  G. Biroli, P. Phillips; hospitality from KITP Santa Barbara; and
  financial support from EPSRC program ``TOPNES'' (EP/I031014/1) and
  EPSRC (EP/G004714/2).  This research was supported in part by the
  National Science Foundation under Grant No. NSF PHY11-25915.}

\end{document}